\documentstyle[prd,preprint,12pt,aps]{revtex}
\begin{document}
\draft
\title{The Topological Structure of the Space-Time Disclination}
\author{Yishi Duan\thanks{E-mail: ysduan@lzu.edu.cn}}
\address{Institute of Theoretical Physics, Lanzhou University, Lanzhou,
730000, P. R. China} 
\author{Sheng Li\thanks{%
Corresponding author. E-mail: lisheng@itp.ac.cn}}
\address{Institute of Theoretical Physics, Academia Sinica, Beijing,
100080, P. R. China}
\date{\today}
\maketitle

\begin{abstract}
The space-time disclination is studied by making use of the decomposition
theory of gauge potential in terms of antisymmetric tensor field and $\phi $%
-mapping method. It is shown that the self-dual and anti-self-dual parts of
the curvature compose the space-time disclinations which are classified in
terms of topological invariants$-$winding number. The projection of
space-time disclination density along an antisymmetric tensor field is
quantized topologically and characterized by Brouwer degree and Hopf index.
\end{abstract}


\section{Introduction}

The topological space-time defects, if they exist, may help to explain some
of the largest-scale structure seen in the universe today. They may have
been found at phase transitions in the early history of the universe,
analogous to those found in some condensed matter systems$-$vortex line in
liquid helium, flux tubes in type-II superconductors, or disclination lines
in liquid crystals.

As a kind of space-time defect, torsion plays an important role in modern
physics. Recently, a lot of works on the spin and torsion have been done by
many physicists.$^{1-11}$ Many of them$^{8-11}$ are focused on the so-called
space-time dislocation by considering the effect of topology of torsion in
the Riemann-Cartan manifold. The consideration partially comes from the
gauge field theory of dislocation and disclination continuum about
deformable material media.$^{12-16}$ These works is trying to clarify the
quantization of the gravitation (i.e. to quantize the Riemann-Cartan
space-time itself). Moreover, they hope to find the dynamical relation
between the stress-energy-momentum tensor and curvature is expressed in
general relativity by Einstein equations.

On the other hand, there exist another kind of space-time defect called
space-time disclination. This kind of defect has the same important meaning
as that of the torsion. Space-time disclination reflects the intrinsic
property of the curved space-time. It is caused by inserting solid angles
into the flat space-time. In Riemann-Cartan geometry, this effect is showed
by the integral of the affine curvature along a closed surface. Duan, Duan
and Zhang$^{12}$ had discussed the disclinations in deformable material
media by applying the gauge field theory and decomposition theory of gauge
potential. In their works, the projection of disclination density along the
gauge parallel vector was found corresponding to a set of isolated
disclinations in the three dimensional sense and being topologically
quantized. The decomposition theory of gauge potential and $\phi $-mapping
method they used are good tools to study the problems of gauge field.

In this paper, we discuss the space-time disclinations in the similar way as
that was used in Duan et al's works on the condensed matter disclination.$%
^{12}$ The decomposition formulas of $SO(4)$ and $SU(2)$ gauge potential are
given in terms of antisymmetric tensor field and unit vector field
respectively. Using this decomposition theory the space-time disclinations
are classified into two classes which are marked by the self-dual and
anti-self-dual parts of the curvature. The projection of space-time
disclination density along the gauge parallel tensor can be written in terms
of winding numbers. Moreover we show that the projection of space-time
disclination density, which is quantized topologically, corresponds to two
groups of isolated disclinations. The positions of the disclination vertices
are determined by the zeroes of the self-dual or anti-self-dual part of an
antisymmetric tensor. And the Hopf index and Brouwer degree classify the
disclinations and characterize the local nature of the space-time
disclinations. For this quantization of the space-time disclinations is
related to the space-time curvature directly as that will be shown in this
paper, it perhaps relates to the quantization of space-time.

This paper is arranged as follows: In section $2$, we discuss the
representation of the space-time disclination. In section $3$, the
decomposition of $SO(4)$ gauge potential is given by considering the
relationship between $SO(4)$ gauge theory and $SU(2)$ gauge theory. At last,
we discuss the topological quantization and the local topological property
of the space-time disclination in section $4$.

\section{The representation of space-time disclination}

Let ${\bf M}$ be a compact, oriented $4$-dimensional Riemannian-Cartan
manifold and $P({\bf M},G,\pi )$ be a principal bundle with structure group
$G=SO(4)$. As it was shown in Ref. 12, the dislocation and disclination
continuum can be described by the reference, deformed and natural states.
For the natural state there is only an anholonomic rectangular coordinate $%
Z^a$ ($a=1,2,3,4$) and 
\begin{equation}
\delta Z^a=e_\mu ^adx^\mu , 
\end{equation}
where $e_\mu ^a$ is vielbein. The metric tensor of the Riemann-Cartan
manifold of natural state is defined by 
\begin{equation}
g_{\mu \nu }=e_\mu ^ae_\nu ^a. 
\end{equation}

We have known that the metric tensor $g_{\mu \nu }$ is invariant under the
local $SO(4)$ transformation of vielbein. The corresponding gauge covariant
derivative 1-form of an antisymmetric tensor field $\phi ^{ab}=-\phi ^{ba}$
on ${\bf M}$ is given as 
\begin{equation}
D\phi ^{ab}=d\phi ^{ab}-\omega ^{ac}\phi ^{cb}-\omega ^{bc}\phi ^{ac}, 
\end{equation}
where $\omega ^{ab}$ is $SO(4)$ spin connection 1-form 
\begin{equation}
\omega ^{ab}=-\omega ^{ba}\quad \quad \omega ^{ab}=\omega _\mu ^{ab}dx^\mu . 
\end{equation}
The affine connection of the Riemann--Cartan space is determined by 
\begin{equation}
\Gamma _{\mu \nu }^\lambda =e^{\lambda a}D_\mu e_\nu ^a. 
\end{equation}
The torsion tensor is the antisymmetric part of $\Gamma _{\mu \nu }$ and is
expressed as 
\begin{equation}
T_{\mu \nu }^\lambda =e^{\lambda a}T_{\mu \nu }^a, 
\end{equation}
where 
\begin{equation}
T_{\mu \nu }^a=\frac 12(D_\mu e_\nu ^a-D_\nu e_\mu ^a). 
\end{equation}
The Riemannian curvature tensor is equivalent to the $SO(4)$ gauge field
strength tensor 2-form $F^{ab}$, which is given by 
\begin{equation}
F^{ab}=d\omega ^{ab}-\omega ^{ac}\wedge \omega ^{cb}\quad \quad F^{ab}=\frac 
12F_{\mu \nu }^{ab}dx^\mu \wedge dx^\nu 
\end{equation}
and relates with the Riemann curvature tensor by 
\begin{equation}
F_{\mu \nu }^{ab}=-R_{\mu \nu \sigma }^\lambda e_\lambda ^ae^{\sigma b}. 
\end{equation}

The dislocation density is defined by 
\begin{equation}
\alpha ^a=T^a\quad \quad \quad T^a=\frac 12T_{\mu \nu }^adx^\mu \wedge
dx^\nu . 
\end{equation}
Analogous to the definition of the 3-dimensional disclination density in the
gauge field theory of condensed matter, we define the space-time
disclination density as 
\begin{equation}
\label{defination-disclination}\theta ^{ab}=\frac 12R_{\mu \nu \sigma
}^\lambda e_\lambda ^ae^{\sigma b}dx^\mu \wedge dx^\nu =-F^{ab}. 
\end{equation}

The size of the space-time disclination can be represented by the means of
the surface integral of the projection of the space-time disclination
density along an antisymmetric tensor field $n^{ab}$ 
\begin{equation}
\label{4w}\Omega =\oint_\Sigma \theta ^{ab}n^{ab}=-\oint_\Sigma F^{ab}n^{ab} 
\end{equation}
where $\Sigma \,$ is a closed surface including the disclinations. The new
quantity $\Omega $ defined by (\ref{4w}) is dimensionless. Using the
so-called $\phi $-mapping method and the decomposition of gauge potential Y.
S. Duan et al have proved that the dislocation flux is quantized in units of
the Planck length.$^{11}$ In this paper we will show the size of the
space-time disclination (\ref{4w}) is topologically quantized and is the sum
of two groups of solid angles representing the size of disclinations.

\section{Decomposition theory of $SO(4)$ spin connection}

In this section we will give the decomposition theory of $SO(4)$ gauge
potential in terms of an antisymmetric tensor field on a compact and
oriented $4$-dimensional manifold, which is the foundation of topological
gauge field theory. It will be seen in the following discussion that this
decomposition of $SO(4)$ gauge potential is expressed by the combination of $%
SU(2)$ gauge potential. Hence we must study the decomposition of $SU(2)$
gauge potential in terms of the sphere bundle on a compact and oriented $4$%
-dimensional manifold firstly.

Let $n$ be an unit $SU(2)$ Lie algebraic vector

\begin{equation}
\label{su2-n}n=n^AI_A\ ,\qquad A=1,2,3\ ; 
\end{equation}
and 
\begin{equation}
\label{2}n^An^A=1, 
\end{equation}
in which $I_A$ is the generator of the group $SU(2)$, which satisfies the
commutation relation 
\begin{equation}
\label{su2-communication}[I_A,I_B]=-\epsilon ^{ABC}I_C. 
\end{equation}
The covariant derivative 1-form of $n$ is given by 
\begin{equation}
\label{su2-Dn}D_{SU(2)}n=dn-[\omega _{SU(2)},n], 
\end{equation}
where $\omega _{SU(2)}$ is the $SU(2)$ gauge potential 1-form: 
\begin{equation}
\label{su2-w}\omega _{SU(2)}=\omega _{SU(2)}^AI_A\ , 
\end{equation}
and 
\begin{equation}
\label{5}\omega _{SU(2)}^A=\omega _{SU(2)\mu }^Adx^\mu \qquad \mu =0,1,2,3. 
\end{equation}
The curvature $2$-form is 
\begin{equation}
\label{su2-F}F_{SU(2)}=F_{SU(2)}^AI_A=d\omega _{SU(2)}-\omega _{SU(2)}\wedge
\omega _{SU(2)}, 
\end{equation}
and 
\begin{eqnarray}
F_{SU(2)}^A&=&\frac 12F_{SU(2)\mu \nu }^Adx^\mu \wedge dx^v\nonumber\\
&=&d\omega _{SU(2)}^A+%
\frac 12\epsilon ^{ABC}\omega _{SU(2)}^B\wedge \omega _{SU(2)}^C. 
\end{eqnarray}
Let $m=m^AI_A,\,l=l^AI_A$ be another two unit $SU(2)$ algebraic vectors
orthornormal to $n$ and to each other, satisfying 
\begin{equation}
n^A=\epsilon ^{ABC}m^Bl^C, 
\end{equation}
It can be proved that the $SU(2)$ gauge potential can be decomposed by $%
n,\,m,\,l$ as

\begin{equation}
\label{su2-wd}\omega _{SU(2)}^A=\epsilon
^{ABC}(dn^Bn^C-D_{SU(2)}n^Bn^C)-n^AA, 
\end{equation}
where 
\begin{equation}
\label{su2-wA}A=dm^Al^A-D_{SU(2)}m^Al^A 
\end{equation}
is a $U(1)$-like gauge potential.

If $n$ is taken as a gauge parallel vector, i.e. 
\begin{equation}
D_{SU(2)}n=0, 
\end{equation}
the $SU(2)$ gauge potential becomes 
\begin{equation}
\label{su2-wp}\omega _{SU(2)}^A=\epsilon ^{ABC}dn^Bn^C-n^AA. 
\end{equation}
Then we can easily get the curvature $2$-form $F_{SU(2)}^A$ as 
\begin{equation}
\label{su2-fp}F_{SU(2)}^A=n^AdA-\frac 12\epsilon ^{ABC}dn^B\wedge dn^C. 
\end{equation}

Now let us consider the $SO(4)$ gauge theory. Let the $4$-dimensional Dirac
matrix $\gamma _a$ ($a=1,2,3,4$) be the basis of the Clifford algebra which
satisfies 
\begin{equation}
\label{clliford-communication}\gamma _a\gamma _b+\gamma _b\gamma _a=2\delta
_{ab}. 
\end{equation}
The antisymmetric tensor field $\phi ^{ab}$ on ${\bf M}$ can be expressed in
the following matrix form 
\begin{equation}
\phi =\frac 12\phi ^{ab}I_{ab}, 
\end{equation}
in which $I_{ab}$ is the generator of the group $SO(4)$%
\begin{equation}
\label{so4-generator}I_{ab}=\frac 14[\gamma _a,\gamma _b]. 
\end{equation}
Similarly, the spin connection $1$-form and curvature $2$-form can be
expressed as 
\begin{equation}
\omega =\frac 12\omega ^{ab}I_{ab},\quad \quad F=\frac 12F^{ab}I_{ab}. 
\end{equation}

It is well known that the spin representation of $SO(4)$ group is hormorphic
to the direct product of the representations of two $SU(2)$ group$^{17}$%
\begin{equation}
so(4)\cong su(2)\otimes su(2). 
\end{equation}
The generators of $SO(4)$ group can be divided into two terms. Each term is
a generator of some $SU(2)$ group. Define 
\begin{eqnarray}
\label{so4-su2-generator}
I_{\pm }^1&=&
\frac 12I_{23}(1\mp \gamma _5)=\frac 14(\gamma _2\gamma _3\pm \gamma
_1\gamma _4); \\
I_{\pm }^2&=&
\frac 12I_{31}(1\mp \gamma _5)=\frac 14(\gamma _3\gamma _1\pm \gamma
_2\gamma _4);
\\ I_{\pm }^3&=&
\frac 12I_{12}(1\mp \gamma _5)=\frac 14(\gamma _1\gamma _2\pm \gamma
_3\gamma _4),
\end{eqnarray}
in which $\gamma _5=\gamma _1\gamma _2\gamma _3\gamma _4$. It can be proved $%
I_{\pm }^A(A=1,2,3)$ satisfy the commutation relation of group $SU(2)_{\pm }$%
\begin{equation}
I_{\pm }^A=-\epsilon ^{ABC}[I_{\pm }^B,I_{\pm }^C], 
\end{equation}
and 
\begin{equation}
[{I_{\pm }^A,I_{\mp }^B]}=0. 
\end{equation}
Therefore $I_{\pm }^A$ are the generators of two different groups $%
SU(2)_{\pm }$, and they are the basis of $SU(2)_{\pm }$ Lie algebraic spaces.

Arbitrary antisymmetric tensor field $\phi ^{ab}$ can be decomposed as 
\begin{equation}
\label{phi-d}\phi ^{ab}=\phi _{+}^{ab}+\phi _{-}^{ab}, 
\end{equation}
where 
\begin{equation}
\phi _{+}^{ab}=\frac 12(\phi ^{ab}+\frac 12\epsilon ^{abcd}\phi ^{cd}) \quad
\quad \phi _{-}^{ab}=\frac 12(\phi ^{ab}-\frac 12\epsilon ^{abcd}\phi ^{cd}) 
\end{equation}
are the self-dual and anti-self-dual parts of $\phi ^{ab}$. Define 
\begin{equation}
\phi _{\pm }^1=\phi ^{23}\pm \phi ^{14},\quad \phi _{\pm }^2=\phi ^{31}\pm
\phi ^{24},\quad \phi _{\pm }^3=\phi ^{12}\pm \phi ^{34}. 
\end{equation}
We can rewrite $\phi $ as 
\begin{equation}
\phi =\frac 12\phi ^{ab}I_{ab}=\phi _{+}+\phi _{-}, 
\end{equation}
and 
\begin{equation}
||\phi _{\pm }||^2=\phi _{\pm }^A\phi _{\pm }^A=\phi _{\pm }^{ab}\phi _{\pm
}^{ab}, 
\end{equation}
in which 
\begin{equation}
\phi _{\pm }=\frac 12\phi (1\mp \gamma _5)=\phi _{\pm }^AI_{\pm }^A. 
\end{equation}
From above discussion, we see that $\phi _{\pm }$ are the $SU(2)_{\pm }$
algebraic vectors. With the similar decomposition as above it can be
verified that 
\begin{equation}
D\phi _{+}+D\phi _{-}=d\phi _{+}-[\omega _{+},\phi _{+}]+d\phi _{-}-[\omega
_{-},\phi _{-}]. 
\end{equation}
For the independent of $I_{+}$ and $I_{-}$ we have 
\begin{equation}
D\phi _{\pm }=d\phi _{\pm }-[\omega _{\pm },\phi _{\pm }]\quad or \quad
D\phi _{\pm }^A=d\phi _{\pm }^A-\epsilon ^{ABC}\omega _{\pm }^C\phi _{\pm
}^B, 
\end{equation}
in which 
\begin{equation}
\omega _{\pm }=\omega _{\pm }^AI_{\pm }^A 
\end{equation}
is the gauge potential of $SU(2)_{\pm }$ gauge field. Similarly curvature $2$%
-form can be decomposed as 
\begin{equation}
F=F_{+}+F_{-}, 
\end{equation}
and 
\begin{equation}
F_{\pm }=d\omega _{\pm }-\omega _{\pm }\wedge \omega _{\pm }. 
\end{equation}
$F_{\pm }=F_{\pm }^AI_{\pm }^A$ are the curvature $2$-forms of $SU(2)_{\pm }$
gauge field.

Now define an antisymmetric tensor 
\begin{equation}
n^{ab}=\frac{\phi _{+}^{ab}}{||\phi _{+}||}+\frac{\phi _{-}^{ab}}{||\phi
_{-}||}, 
\end{equation}
i.e. 
\begin{equation}
n_{+}^{ab}=\frac{\phi _{+}^{ab}}{||\phi _{+}||},\quad \quad \quad \quad
\quad n_{-}^{ab}=\frac{\phi _{-}^{ab}}{||\phi _{-}||}. 
\end{equation}
Then 
\begin{equation}
n=\frac 12n^{ab}I_{ab}=n_{+}+n_{-} 
\end{equation}
and 
\begin{equation}
\label{n-d}n_{\pm }=\frac 1{2||\phi _{\pm }||}\phi (1\mp \gamma _5)=n_{\pm
}^AI_{\pm }^A. 
\end{equation}
It naturally guarantees the constraint 
\begin{equation}
n_{\pm }^An_{\pm }^A=1. 
\end{equation}
i.e. $n_{\pm }^A$ are the $SU(2)_{\pm }$ lie algebraic unit vectors. If $n$
is taken as a gauge parallel tensor field 
\begin{equation}
\label{phi-p}Dn=0, 
\end{equation}
we have 
\begin{equation}
Dn_{\pm }=dn_{\pm }-[\omega _{\pm },n_{\pm }]=0. 
\end{equation}

From the decomposition formula of $SU(2)$ gauge potential (\ref{su2-fp}) we
get 
\begin{equation}
\label{so4-Fd}F_{\pm }^A=n_{\pm }^AdA_{\pm }-\frac 12\epsilon ^{ABC}dn_{\pm
}^B\wedge dn_{\pm }^C 
\end{equation}
This formula is very useful in the following discussing of space-time
disclinations.

\section{The topological quantization and the local topological structure of
the space-time disclination}

It is easy to see that 
\begin{equation}
F^{ab}n^{ab}=F_{+}^{ab}n_{+}^{ab}+F_{-}^{ab}n_{-}^{ab} 
\end{equation}
and 
\begin{equation}
F_{\pm }^An_{\pm }^A=F_{\pm }^{ab}n_{\pm }^{ab}. 
\end{equation}
By considering (\ref{so4-Fd}) we have 
\begin{equation}
F_{\pm }^An_{\pm }^A=dA_{\pm }-\frac 12\epsilon ^{ABC}n_{\pm }^Adn_{\pm
}^B\wedge dn_{\pm }^C. 
\end{equation}
Define 
\begin{equation}
\label{W-d}\Omega _{\pm }=\oint_\Sigma (\frac 12\epsilon ^{ABC}n_{\pm
}^Adn_{\pm }^B\wedge dn_{\pm }^C-dA_{\pm }). 
\end{equation}
$\Omega _{\pm }$ represent the components of space-time disclination
corresponding to different gauge group $SU(2)_{\pm }$. The space-time
disclination can be expressed as 
\begin{equation}
\Omega =\Omega _{+}+\Omega _{-} 
\end{equation}
If the surface $\Sigma $ is closed, the second term of the right side of (%
\ref{W-d}) contribute nothing to $\Omega _{\pm }$, i.e. 
\begin{equation}
\Omega _{\pm }=\oint_\Sigma \frac 12\epsilon ^{ABC}n_{\pm }^Adn_{\pm
}^B\wedge dn_{\pm }^C. 
\end{equation}
To study the relationship between the solid angle $\Omega _{\pm }$ and the
local properties of the disclinations inside the closed surface $\Sigma $,
using the Stoke's formula, $\Omega _{\pm }$ can be expressed as 
\begin{equation}
\label{disclinations}\Omega _{\pm }=\int_V\frac 12\epsilon ^{ABC}dn_{\pm
}^A\wedge dn_{\pm }^B\wedge dn_{\pm }^C, 
\end{equation}
in which $\partial V=\Sigma $. Let us choose coordinates $y=(u^1,u^2,u^3,v)$ on 
$M$ such that $u=(u^1,u^2,u^3)$ are intrinsic coordinate on $V$. For the
coordinate component $v$ does not belong to $V$. Then 
\begin{eqnarray}
\Omega _{\pm } & =&\int_V
\frac 12\epsilon ^{ABC}\partial _in_{\pm }^A\partial _jn_{\pm }^B\partial
_kn_{\pm }^Cdu^i\wedge du^j\wedge du^k \nonumber\\
  & =&\int_V\frac 12\epsilon
^{ijk}\epsilon ^{ABC}\partial _in_{\pm }^A\partial _jn_{\pm }^B\partial
_kn_{\pm }^Cd^3u
\end{eqnarray}
where $i,j,k=1,2,3$ and $\partial _i=\partial /\partial u^i$. Define solid
angle densities as 
\begin{equation}
\label{solid-angle}\rho _{\pm }=\frac 12\epsilon ^{ijk}\epsilon
^{ABC}\partial _in_{\pm }^A\partial _jn_{\pm }^B\partial _kn_{\pm }^C. 
\end{equation}
Then we get 
\begin{equation}
\Omega _{\pm }=\int_V\rho _{\pm }d^3u. 
\end{equation}

For the equation (\ref{n-d}), the unit vectors $n_{\pm }^A(x)$ can expressed
as follows: 
\begin{equation}
n_{\pm }^A=\frac{\phi _{\pm }^A}{||\phi _{\pm }||} 
\end{equation}
Hence 
\begin{equation}
dn_{\pm }^A=\frac 1{||\phi _{\pm }||}d\phi _{\pm }^A+\phi _{\pm }^Ad(\frac 1{%
||\phi _{\pm }||}), 
\end{equation}
and 
\begin{equation}
\frac \partial {\partial \phi _{\pm }^A}(\frac 1{||\phi _{\pm }||})=-\frac{%
\phi _{\pm }^A}{||\phi _{\pm }||^3}. 
\end{equation}
Substituting above equations into the solid angle density (\ref{solid-angle}%
) we obtain 
\begin{eqnarray}
\rho & =&
\frac 12\epsilon ^{ABC}\epsilon ^{ijk}\partial _i(n_{\pm }^A\partial
_jn_{\pm }^B\partial _kn_{\pm }^C) \nonumber \\
& =&\frac 12\epsilon ^{ABC}\epsilon ^{ijk}\partial _i\frac{\phi _{\pm }^A}{||\phi _{\pm }||^3}\partial _j\phi _{\pm }^B\partial _k\phi _{\pm }^C
\nonumber \\
& =&-\frac 12\epsilon ^{ABC}\epsilon ^{ijk}\frac \partial {\partial \phi _{\pm
}^D}\frac \partial {\partial \phi _{\pm }^A}(\frac 1{||\phi _{\pm }||}%
)\partial _i\phi _{\pm }^D\partial _j\phi _{\pm }^B\partial _k\phi _{\pm
}^C. 
\end{eqnarray}
Define the Jacobian $J(\frac{\phi _{\pm }}u)$ as 
\begin{equation}
\epsilon ^{ABC}J(\frac{\phi _{\pm }}u)=\epsilon ^{ijk}\partial _i\phi _{\pm
}^A\partial _j\phi _{\pm }^B\partial _k\phi _{\pm }^C. 
\end{equation}
By making use of the Laplacian relation in $\phi $-space 
\begin{equation}
\partial _A\partial _A\frac 1{||\phi _{\pm }||}=-4\pi \delta ^3(\phi _{\pm
}),\quad \partial _A=\frac \partial {\partial \phi _{\pm }^A}, 
\end{equation}
we can write the density of the solid angle as the $\delta $-like expression 
\begin{equation}
\label{density-1}\rho _{\pm }=4\pi \delta ^3(\phi _{\pm })J(\frac{\phi _{\pm
}}u) 
\end{equation}
and 
\begin{equation}
\Omega _{\pm }=\int_V4\pi \delta ^3(\phi _{\pm })J(\frac{\phi _{\pm }}u%
)d^3u. 
\end{equation}
It obvious that $\rho _{\pm }$ are non-zero only when $\phi _{\pm }=0$.

Suppose that $\phi _{\pm }^A(x)$ ($A=1,2,3)$ possess $K_{\pm }$ isolated
zeros, according to the deduction of Ref. 18 and the implicit
function theorem,$^19$ the solutions of $\phi _{\pm
}(u^1,u^2,u^3,v)=0$ can be expressed in terms of $u=(u^1,u^2,u^3)$ as 
\begin{equation}
u^i=z_{\pm }^i(v),\quad \quad \quad \quad \quad i=1,2,3 
\end{equation}
and 
\begin{equation}
\phi _{\pm }^A(z_l^1(v),z_l^2(v),z_l^3(v),v)\equiv 0, 
\end{equation}
where the subscript $l=1,2,\cdots ,K_{\pm }$ represents the $l$th zero of $%
\phi _{\pm }^A$, i.e. 
\begin{equation}
\phi _{\pm }^A(z_{\pm l}^i)=0,\quad \quad \quad l=1,2,\cdots ,K_{\pm };\quad
A=1,2,3. 
\end{equation}
It is easy to get the following formula from the ordinary theory of $\delta $%
-function that 
\begin{equation}
\delta ^3(\phi _{\pm })J(\frac{\phi _{\pm }}u)=\sum_{i=1}^{K_{\pm }}\beta
_{\pm l}\eta _{\pm l}\delta ^3(u-z_{\pm l}) 
\end{equation}
in which 
\begin{equation}
\eta _{\pm l}=signJ(\frac{\phi _{\pm }}u)|_{x=z_{\pm l}}=\pm 1, 
\end{equation}
is the Brouwer degree of $\phi $-mapping and $\beta _{\pm l}$ are positive
integers called the Hopf index of map $\phi _{\pm }$ which means while the
point $x$ covers the region neighboring the zero $x=z_{\pm l}$ once, $\phi
_{\pm }$ covers the corresponding region $\beta _{\pm l}$ times. Therefore
the slid angle density becomes 
\begin{equation}
\label{density-d}\rho _{\pm }=4\pi \sum_{l=1}^{K_{\pm }}\beta _{\pm l}\eta
_{\pm l}\delta ^3(u-z_{\pm l}) 
\end{equation}
and 
\begin{equation}
\Omega _{\pm }=4\pi \sum_{l=1}^{K_{\pm }}\beta _{\pm l}\eta _{\pm
l}\int_V\delta ^3(u-z_{\pm l})d^3u=4\pi \sum_{l=1}^{K_{\pm }}\beta _{\pm
l}\eta _{\pm l} 
\end{equation}
Therefore the space-time disclination is 
\begin{equation}
\Omega =4\pi \sum_{l=1}^{K_{+}}\beta _{+l}\eta _{+l}+4\pi
\sum_{l=1}^{K_{-}}\beta _{-l}\eta _{-l} 
\end{equation}

We find that (\ref{density-d}) is the exact density of a system of $K_{+}$
and $K_{-}$ classical point-like objects with ``charge'' $\beta _{+l}\eta
_{+l}$ and $\beta _{-l}\eta _{-l}$ in space-time, i.e. the topological
structure of disclinations formally corresponds to a point-like system.
These point objects may be called disclination points as in nematic crystals.%
$^{20}$ In Ref. 21, it was shown that the existence of disclination points
is related to a kind of broken symmetries. The dislocations and
disclinations appear as singularities of distortions of an order parameter.$%
^{22}$ In our paper, the disclination points are identified with the
isolated zero points of vector field $\phi _{\pm }^A(x)$. From (\ref
{density-1}) we know that these singularities are those of the disclination
density as well.

On another hand, the winding number $W_{\pm}$ of the surface $\Sigma $ and of
the mapping $\phi_{\pm }$ is defined as$^{23}$
\begin{equation}
\label{winding}W_{\pm }=\oint_\Sigma \frac 1{8\pi }\epsilon ^{ABC}\frac{\phi
_{\pm }^A}{||\phi _{\pm }||^3}d\phi _{\pm }^B\wedge d\phi _{\pm }^C 
\end{equation}
which is equal to the number of times $\Sigma $ encloses (or, wraps around)
the point $\phi _{\pm }=0$. Hence, the space-time disclinations is quantized
by the winding numbers 
\begin{equation}
\label{windingnumber}\Omega _{\pm }=4\pi W_{\pm }. 
\end{equation}
The winding number $W_{\pm }$ of the surface $\Sigma $ can be interpreted
or, indeed, defined as the degree of the mappings $\phi _{\pm }$ onto $%
\Sigma $. By (\ref{disclinations}) and (\ref{density-1}) we have
\begin{eqnarray}
\Omega _{\pm } & =&4\pi \int_V\delta (\phi _{\pm })J_{\pm }(
\frac{\phi _{\pm }}u)d^3u \nonumber\\
 & =&4\pi \deg \phi _{\pm }\int_{\phi _{\pm
}(V)}\delta (\phi _{\pm })d^3\phi _{\pm } \nonumber\\
& =&4\pi \deg \phi _{\pm }
\end{eqnarray}
where $\deg \phi _{\pm }$ are the degrees of map $\phi _{\pm }:V\rightarrow
\phi _{\pm }(V).$ Compared above equation with (\ref{windingnumber}), it
shows the degrees of map $\phi _{\pm }:V\rightarrow \phi _{\pm }(V)$ is just
the winding number $W_{\pm }$ of surface $\Sigma $ and map $\phi _{\pm }$,
i.e. 
\begin{equation}
\deg \phi _{\pm }=W_{\pm }(\Sigma ,\phi _{\pm }). 
\end{equation}
Then the space-time disclination is 
\begin{equation}
\Omega =4\pi (\deg \phi _{+}+\deg \phi _{-})=4\pi (W_{+}+W_{-}). 
\end{equation}
Divide $V$ by 
\begin{equation}
V=\sum_{l=1}^{K_{\pm }}V_{\pm l} 
\end{equation}
and $V_{\pm l}$ includes only one zero $z_{\pm l}$ of $\phi _{\pm }$, i.e. $%
z_{\pm l}\in V_{\pm l}$. The winding number $W_{\pm }$ of the surface $%
\Sigma _{\pm l}=\partial V_{\pm l}$ and the mapping $\phi _{\pm }$ is
defined as 
\begin{equation}
W_{\pm l}=\oint_{\Sigma _{\pm l}}\frac 1{8\pi }\epsilon ^{ABC}\frac{\phi
_{\pm }^A}{||\phi _{\pm }||^3}d\phi _{\pm }^B\wedge d\phi _{\pm }^C, 
\end{equation}
which is equal to the number of times $\Sigma _{\pm l}$ encloses (or, wraps
around) the point $\phi _{\pm }=0$. It is easy to see that 
\begin{equation}
W_{\pm }=\sum_{l=1}^{K_{\pm }}W_{\pm l} 
\end{equation}
and 
\begin{equation}
|W_{\pm l}|=\beta _{\pm l}. 
\end{equation}
Then 
\begin{equation}
\Omega =4\pi \sum_{l=1}^{K_{+}}W_{+l}+4\pi \sum_{l=1}^{K_{-}}W_{-l}. 
\end{equation}
According that is proved in Ref. 24, the increment 
\begin{equation}
\frac 12\epsilon ^{ABC}\frac{\phi _{\pm }^A}{||\phi _{\pm }||^3}d\phi _{\pm
}^B\wedge d\phi _{\pm }^C=\epsilon ^{ABC}\epsilon _{ijk}\frac{\phi _{\pm }^A%
}{||\phi _{\pm }||^3}\partial _i\phi _{\pm }^B\partial _j\phi _{\pm }^CdS^k 
\end{equation}
may be regarded as the solid angle on $\phi _{\pm }(\Sigma _{\pm l})$
subtended by the surface element $dS$ on $\Sigma _{\pm l}$. Now it is easy
to see why the space-time disclination is caused by inserting solid angles
into the flat space-time.

Also we can write 
\begin{equation}
\Omega =4\pi (N_{+}^{+}+N_{-}^{+})-4\pi (N_{+}^{-}+N_{-}^{-}) 
\end{equation}
in which $N_{+}^{\pm }$ are the sums of the Hopf indexes with respect to $%
\eta _{+}=\pm 1$ and $N_{-}^{\pm }$ are the sums of the Hopf indexes with
respect to $\eta _{-}=\pm 1$. We see that while $x$ covers $V$ once, $\phi
_{+}^A$ must cover $\phi _{+}(V)$ $N_{+}^{+}$ times with $\eta =1$ or $%
N_{+}^{-}$ times with $\eta =-1$ while $\phi _{-}^A$ must cover $\phi
_{-}(V) $ $N_{-}^{+}$ times with $\eta =1$ or $N_{-}^{-}$ times with $\eta
=-1.$ Therefore the topological disclinations are distinguished by the sign
of the Jacobian $J(\phi _{\pm }/u)$ at the zero point $x^\mu =z_{\pm l}^\mu $%
. Consequently, the global topological property of disclinations is
characterized by the Brouwer degrees and Hopf indices of each disclination
point, which label the local structure of disclinations.

In this paper, we have studied the topological structure, global and local
properties of space-time disclinations. The disclinations we discussed are
the general $4$-dimensional case, which is characterized by the so-called
topological points. The decomposition of gauge potential, introduction of
the gauge parallel tensor field $n^{ab}(x)$ $(a,b=1,2,3,4)$ and the $\phi $%
-mapping method play important roles in establishing our theory. The
disclination points are the zeros of the self-dual or anti-self-dual parts
of the tensor field $\phi ^{ab}(x)$ and the topological characteristics of
the space-time disclinations are determined by the winding numbers which can
be represented by the Brouwer degrees and Hopf indices at the disclination
points. From the definition of the space-time disclination (\ref
{defination-disclination}), we know that the space-time disclination is
expressed exactly by the curvature of space-time. Therefore the quantization
of the space-time disclination has close relationship with the quantization
of space-time.

\acknowledgments
This research was supported by National Natural Science Foundation of P. R.
China.

\newpage\ \noindent {\bf Reference}

\begin{itemize}
\item[$^1$]  R. Jha, Int. J. Mod. Phys. {\bf A9, 3959} (1994);

\item[$^2$]  L. L. Smalley and J. P. Krisch, Class. Quantum Grav. {\bf 11},
2375 (1994);

\item[$^3$]  R. Hammond, Gen. Rel. Grav. {\bf 26}, 1107 (1994);

\item[$^4$]  C. Sivaram and L. C. Garica De Andrade, Astro. Space Sci. {\bf %
201}, 131 (1993);

\item[$^5$]  J. K\'ann\'ar, Gen. Rel. Grav. {\bf 27}, 23 (1995);

\item[$^6$]  C. M. Zhang, Int. J. Mod. Phys. {\bf A8}, 5095 (1993);

\item[$^7$]  V. de Sabbata and Y. Xin, Int. J. Mod. Phys. {\bf A10}, 3663
(1995);

\item[$^8$]  V. de Sabbata, IL Nuovo Cimemto. {\bf A107}, 363 (1994);

\item[$^9$]  S. Luo, Int. J. Theor. Phys. {\bf 34}, 2009 (1995);

\item[$^{10}$]  D.K. Ross, Int. J. Theor. Phys. {\bf 28}, 1333 (1989);

\item[$^{11}$]  Y. S. Duan and S. L. Zhang and S. S. Feng, J. Math. Phys. {\bf %
35}, 1{\bf \ } (1994); Y. S. Duan, G. H. Yang and Y. Jiang, Gen. Rel. Grav. 
{\bf 29}, 715 (1997); Y. S. Duan, G. H.Yang and Y. Jiang, Helv. Phys. Acta. 
{\bf 70}, 565 (1997);

\item[$^{12}$]  Y. S. Duan, and Z. P. Duan, Int. J. Engng. Sci. {\bf 24}, 513
(1986); Y. S. Duan, and S. L. Zhang, Int. J. Engng. Sci.{\bf \ 28}, 689{\bf %
\ }(1990){\bf ; 29}, 153 (1991);{\bf \ 30}, 153 (1992);

\item[$^{13}$]  D. G. Edelen, Int. J. Engng. Sci. {\bf 18}, 1095 (1980); A.
Kadic and D. G. B. Edelen, Int. J. Engng. Sci. {\bf 20}, 443 (1982); in {\it %
A gauge field theory of Dislocations and Discliantions}, Lecture Notes in
Physics. No. 174. Springer, Berlin (1983);

\item[$^{14}$]  D. G. B. Edelen and D. C. Lagoudas, in {\it Gauge Theory and
Defects in Solids.} North-Holland, Amsterdam (1988);

\item[$^{15}$]  N. D. Mermin, Rev. Mod. Phys. {\bf 51}, 591 (1979); L. Michel,
Rev. Mod. Phys. {\bf 52}, 617 (1980); H. R. Trebin, Adv. Phys. {\bf 31}, 195
(1982);

\item[$^{16}$]  B. K. D. Gairola, in {\it Dislocations in Solids} (Edited by F.
R. N. Nabarro), Vol. 1, North-Holland, Amsterdam (1979);

\item[$^{17}$]  H. Boerner, in {\bf {\it Representation of Groups} }%
North-Holland Publishing Company, 1963{\bf ;}

\item[$^{18}$]  Y. S. Duan and J. C. Liu, in {\it Proceedings of the Johns
Hopkins Workshop on Current Problems in Particle Theory, }{\bf 11 }(1987)
183;

\item[$^{19}$]  \'Edouard Goursat, in {\it A Course in Mathematical Analysis,}
Vol. I (translated by Earle Raymond Hedrick), 1904;

\item[$^{20}$]  M. Kleman, in {\it Dislocations in Solids} (Edited by F. R. N.
Nabarro), Vol. 5, North-Holland, Amsterdam (1979);

\item[$^{21}$]  P. G. de Gennes, C.\ R. Acad. Sci. (Paris) {\bf 275}, 319
(1972);

\item[$^{22}$]  J. Friedel, in {\it Dislocations in Solids} (Edited by F. R. N.
Nabarro), Vol. 5, North-Holland, Amsterdam (1979);

\item[$^{23}$]  Victor Guillemin and Alan Pollack, in {\it Differential Topology%
}, Prentice-Hall, Inc., Englewood Cliffs, New Jersey, 1974;

\item[$^{24}$]  D. Hestenes and G. Sobczyk, in {\it Clifford Algebra to
Geometric Calculus} Reidel, Dordrecht 1984.
\end{itemize}

\end{document}